# Determination of the acceleration due to gravity by studying magnet's motion through a conducting pipe

Sanjoy Kumar Pal[1,4], Soumen Sarkar[2,4], and Pradipta Panchadhyayee[3,4*]

[1]Anandapur H.S. School, Anandapur, Paschim Medinipur, West Bengal, India
[2]Karui P.C. High School, Hooghly, West Bengal, India
[3]Department of Physics (UG & PG), Prabhat Kumar College, Contai, Purba Medinipur, India
[4]Institute of Astronomy Space and Earth Science Kolkata -700054, W. B., India
[*]E-mail: ppcontai@gmail.com

## Abstract

We determine the acceleration due to gravity ($g$) in a novel way using a magnetic sensor and video analysis technique of a smartphone. The same applications are used to measure the terminal velocity of a magnet falling through a conducting pipe and the magnetic moment of the magnet from its torsional oscillations. This experiment would appear to be intriguing, as it combines elements of magnetism, terminal velocity, and electromagnetic damping to determine $g$.

## Introduction

The integration of smartphone sensors into introductory and first-year university physics courses has become increasingly popular in recent years, opening up exciting opportunities for hands-on experimentation and real-world applications of physics concepts. Besides, smartphone-based physics teaching has indeed gained significant attention and popularity in recent years. This approach leverages the capabilities of smartphones to enhance the learning and teaching of physics concepts in various ways [1-3].

Determining the acceleration due to gravity ($g$) is a fundamental experiment in introductory physics classes. The diverse experiments in this aspect help students understand the concept of gravity and how it affects objects on Earth. Several methods using smartphone sensors [4-11] have already been reported to measure $g$. The common sensors in a standard smartphone used for experimentation are the magnetic field sensor, ambient light sensor, sound sensor, and accelerometer. These sensors can be pretty handy for conducting physics experiments in a more accessible and engaging way. The smartphone accelerometer finds its application in measuring $g$ through a thorough investigation of the free fall of an object [4], even under the consideration of friction. The ambient light sensor is used to measure $g$ by recording the oscillatory motion of a pendulum and thereby measuring the time period [5,6]. Analyzing the free-falling motion by employing a 'multitasking method' and using the voice memo app installed on a smartphone [7], the acceleration due to gravity is measured. An experimental method is proposed to measure $g$ by measuring the period of oscillation of a simple pendulum using the proximity sensor integrated into a smartphone [8]. The determination of $g$ is also performed with a bunch of smartphone-based experiments [9]. In another experiment [10], the use of the acoustic stopwatch sensor of the smartphone is explored to determine $g$ accurately by recording the time of free-falling motion. Recently, an experiment [11] is reported to determine $g$ by adopting the 'SONAR' principle with the use of a smartphone for double purposes, i.e. to generate sound waves of a particular frequency and record the result of the sound wave resonance.

In this work, we have measured the acceleration due to gravity by measuring the magnetic moments of four neodymium ball magnets of different sizes as well as their terminal velocities during their motion through a conducting pipe. Our study has two parts. In the first part, we treat the ball magnets as the bobs of a torsional pendulum. The corresponding magnetic moments are measured by studying torsional oscillations with the help of the magnetic sensor of a smartphone. We have shown the variation of magnetic moments in size. Further, the magnetisation of the material of the ball magnets is computed. In the second part, we have allowed the magnets to fall under gravity through a non-ferromagnetic conducting pipe. In a recent work [12], the magnetic moment of a neodymium cylindrical magnet falling through a non-ferromagnetic conducting pipe has been determined. As is shown in Ref. [12], the cylindrical magnets are found to attain the respective terminal velocities under the simultaneous action of the force of gravity and electromagnetic damping. We have measured the terminal velocities using the magnetic sensor of the smartphone. The precision in measurement is enhanced by the processes of tracking and analyzing videos of the measurement setup. On measuring the respective terminal velocities and magnetic moments for each ball, the acceleration due to gravity is calculated and averaged to obtain the final value of $g$.

## Theoretical framework

In the first part to compute the magnetic moment of a magnet, a neodymium ball magnet is suspended by a thread of negligible torsional rigidity from a rigid support and used as a bob of a pendulum. The ball magnet is oriented stable along the north-south (*N*-*S*) direction due to earth's magnetic field. It is well known that it will follow simple harmonic motion if it is twisted a little. The formula for the time period ($T$) is $T = 2\pi\sqrt{\frac{I}{mB_H+c}} \approx 2\pi\sqrt{\frac{I}{mB_H}} = 2\pi\sqrt{\frac{2Mr^2}{5mB_H}}$, (1) where $T$ is the time period of torsional oscillations, $B_H$ is the horizontal component of Earth's magnetic field, and $c$ is torsional rigidity (negligible) of the thread. Here $m$, $M$, and $r$ are the magnetic moment, mass, and radius of the magnet, respectively. $I$ is the moment of inertia of the ball magnet executing a torsional motion, which is equal to $\frac{2}{5}Mr^2$ because the ball magnet is treated as a solid sphere. Measuring the time period accurately we have computed the magnetic moment of the magnet using the known physical parameters of the ball magnet and the earth's magnetic field.

Hence, we determine the value of the intensity of magnetization ($I_M$) of the material of the magnets from the equation given below:

$I_M = m / V$

Therefore, $m = I_M.V = I_M (M/\rho) = (I_M/\rho) M$, (2)

where $\rho$ is the density of the material of the magnet and $V$ is the volume of the magnet.

In the second part we have determined the terminal velocity of the same ball magnet when it undergoes falling under gravity through a conducting pipe. When a magnet is dropped and allowed to move through a conducting pipe, it creates a changing magnetic field in the pipe. As the magnetic field inside the pipe changes, it induces eddy currents which will flow in a way that tries to counteract the motion of the magnet. The eddy currents above and below the falling magnet will encircle the conducting pipe in opposite directions. This is because they are trying to oppose the change in the magnetic flux produced by the falling magnet. The eddy currents act like small electromagnets. When the *S*-pole of the falling magnet is leading the fall, the eddy currents around the *N*-pole will create a magnetic field that attracts the *N*-pole of the magnet from the top of the pipe and repels the *S*-pole from the bottom of the pipe. This electromagnetic interaction between the eddy currents and the falling magnet creates a resistive force that opposes the free-falling motion of the magnet and finally slows down the magnet's fall. As a result, the magnet in falling motion faces a high viscous medium created in the pipe. For this kind of motion of the ball magnet it is obvious that, after a considerable time, the damping force ($F_{em}$) eventually becomes equal to its weight. In this condition, no net force acts on the falling magnet. So, the electromagnetic force, ($F_{em}$) experienced by the ball magnet is proportional to the terminal velocity ($v_T$) attained by the magnet when we neglect the viscous force, the buoyant force, and hydrodynamic resistance (proportional to the square of the instantaneous velocity) of air medium.

$$F_{em} = kv_T. \quad (3)$$

Here the expression of the proportionality constant $k$ is

$$k = \left(\frac{15}{1024}\right)\mu_0^2 m^2 \sigma \left(\frac{1}{a^3} - \frac{1}{b^3}\right), \quad (4)$$

where $\sigma$ is the conductivity of the material of the pipe. Here, $a$ and $b$ are the inner and outer radii of the pipe, $\mu_0$ is the permittivity of free space.

As stated before, when the damping force becomes equal to the weight of the falling magnet, the velocity of the magnet reaches its terminal velocity ($v_T$). Now we can write [12]

$$kv_T = Mg, \quad (5)$$

So, $\left(\frac{15}{1024}\right)\mu_0^2 m^2 \sigma \left(\frac{1}{a^3} - \frac{1}{b^3}\right) v_T = Mg,$

$\therefore m^2 v_T = \frac{Mg}{U}$, where the constant, $U = \left(\frac{15}{1024}\right)\mu_0^2 \sigma \left(\frac{1}{a^3} - \frac{1}{b^3}\right)$.

Hence, $g = \frac{m^2 v_T U}{M}$. (6)

Using the above expression (6) the value of $g$ is calculated where the constant, $U$, only depends upon the system parameters.

**Experimental Results**

As mentioned before, we have performed the experiment for measuring g in two steps. Four ball magnets are taken for this purpose. We have measured the radius of each magnetic ball by digital slide callipers of vernier constant 0.01 mm and the respective masses by a digital scale balance.

Table 1: Masses and Diameters of four ball magnets

| Ball magnet No. | 1 | 2 | 3 | 4 |
|---|---|---|---|---|
| Mass (g) | 0.260 | 0.501 | 0.810 | 1.911 |
| Diameter (mm) | 4.03 | 5.03 | 5.88 | 7.81 |

**I.** *Measurement of the magnetic moments of the ball magnets*

With a view to measuring the magnetic moment of a ball magnet, we have made a torsional pendulum (Fig. 1a) using the magnet and measured the time period of the torsional oscillations. First, to identify the pole of a neodymium ball magnet (a strong but light-weight magnet) is placed on the floor. It is observed that it immediately aligns itself parallel to Earth's magnetic field. To make it sure, we put a black mark on the north pole of the ball magnet by a marker and test the proper alignment of magnet parallel to Earth's magnetic field using magnetometer in phyphox application installed in a smartphone for several times. By hanging the ball magnet with a thread a torsional pendulum is prepared. Now the Magnetometer in phyphox app shows the waveforms corresponding to the periodic variation in the magnetic field with time due to the torsional oscillations of the magnet (See Fig. 1b). Time period has been measured from the time difference related to the two peaks with an interval of $n$ peaks of the magnetic field variation and computed the average time period from a number of observations. For accuracy, the smartphone is kept at around 20 cm apart from the pendulum so that inbuilt magnets (speaker, magsafe charging of iPhone) of the iPhone do not influence the oscillation of the ball magnet. For precaution, we have ensured that there was no ferromagnetic material near the setup. The earth's magnetic field is also measured with the help of the magnetometer in the phyphox app. Following the same procedure of measuring the time periods, we have calculated the magnetic moments (see Table 2) for four ball magnets of different sizes using the values of all required quantities, as shown in Eq. (1).

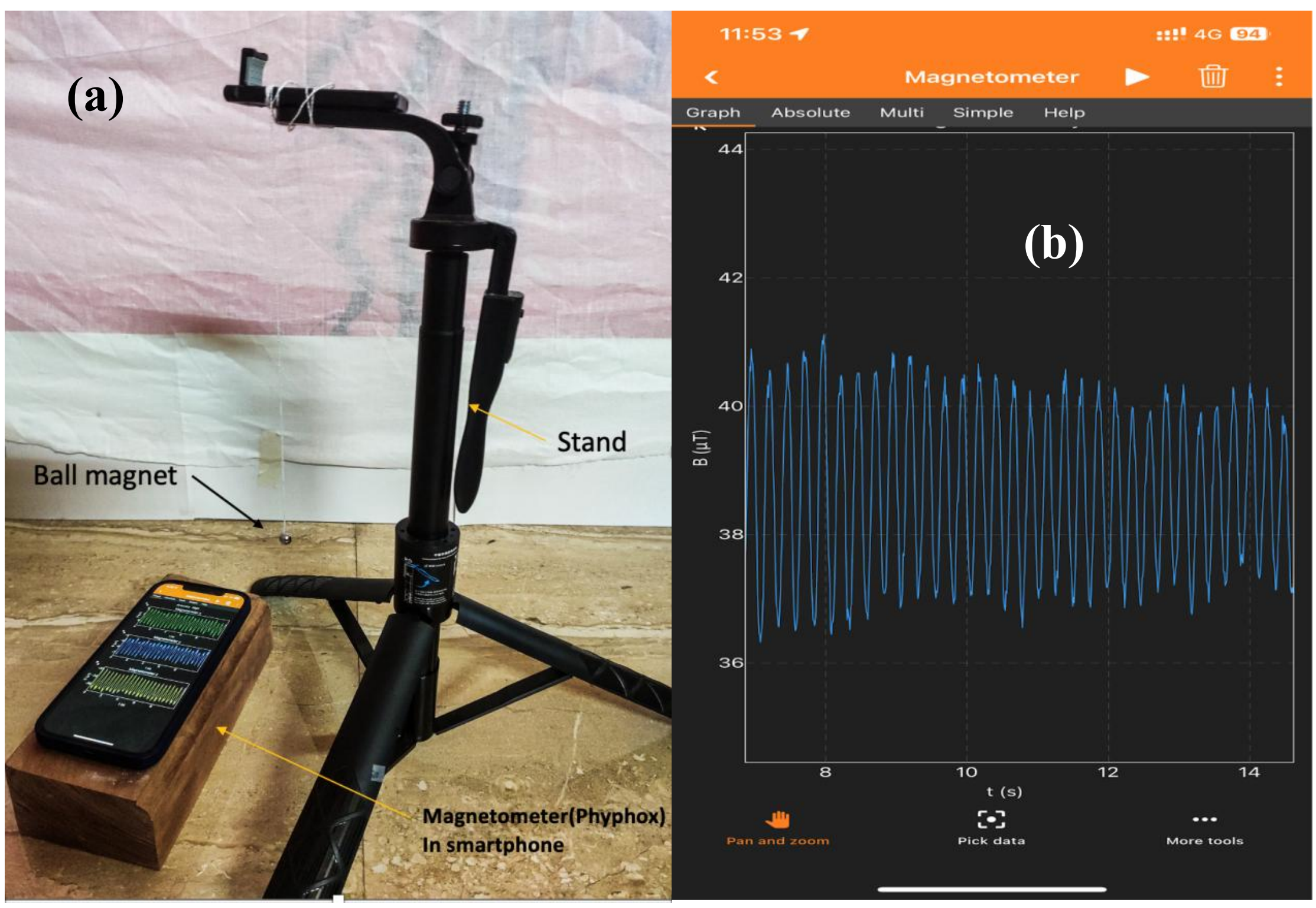

Fig. 1: (a) Experimental setup for studying torsional oscillation of a magnet. (b) The waveforms of magnetic field variation with time are seen in the Magnetometer in phyphox app.

Table 2: Calculation of the magnetic moments for four ball magnets studying their torsional oscillations

| Ball magnet No. | Moment of inertia, $I$ (kg-m$^2$) x10$^{-10}$ | $B_H$ (μT) | No. of oscillations | Total time taken (s) | Time period, $T$ (s) | Mag. moment, $m$ (A-m$^2$) | Mean value of $m$ (A-m$^2$) |
|---|---|---|---|---|---|---|---|
| 1 | 4.222 | 35.03 | 34 | 4.4 | 0.1294 | 0.0283 | 0.0284 ± 0.0004± 0.00004 |
| | | | 31 | 4.0 | 0.1290 | 0.0286 | |
| | | | 48 | 6.2 | 0.1292 | 0.0285 | |
| | | | 44 | 5.7 | 0.1295 | 0.0283 | |
| | | | 41 | 5.3 | 0.1293 | 0.0285 | |
| 2 | 12.676 | 35.03 | 60 | 10.5 | 0.1750 | 0.0466 | 0.0466 ± 0.0005± 0.0001 |
| | | | 52 | 9.1 | 0.1750 | 0.0466 | |
| | | | 63 | 11.0 | 0.1746 | 0.0468 | |
| | | | 55 | 9.6 | 0.1745 | 0.0468 | |
| | | | 65 | 11.4 | 0.1754 | 0.0464 | |
| 3 | 28.005 | 35.03 | 74 | 15.1 | 0.2041 | 0.0757 | 0.0760 ± 0.0007± 0.0002 |
| | | | 62 | 12.6 | 0.2032 | 0.0763 | |
| | | | 81 | 16.5 | 0.2037 | 0.0760 | |
| | | | 67 | 13.6 | 0.2030 | 0.0765 | |
| | | | 71 | 14.5 | 0.2042 | 0.0756 | |
| 4 | 116.563 | 35.03 | 111 | 29.8 | 0.2685 | 0.1821 | 0.1821 ± 0.0011± 0.0002 |
| | | | 106 | 28.4 | 0.2679 | 0.1828 | |
| | | | 83 | 22.3 | 0.2687 | 0.1818 | |
| | | | 102 | 27.4 | 0.2686 | 0.1819 | |
| | | | 111 | 29.8 | 0.2685 | 0.1821 | |

A graph (mass ($M$) versus magnetic moment ($m$)) is plotted to calculate the slope of the graph. It is nothing but a straight line whose slope indicates the value of ($I_M/\rho$) (based on the expression (2)), which is equal to 94.281 A-m$^2$kg$^{-1}$. Measuring the mass and the radius of the individual ball magnets (see Table 1), we have found the average value of $\rho$ as 7593.16 kg-m$^3$, which leads to the computation of the intensity of magnetization ($I_M$) of the material of the magnets. The value of $I_M$ is calculated as (7.2±0.1) x 10$^7$ A-m$^{-1}$.

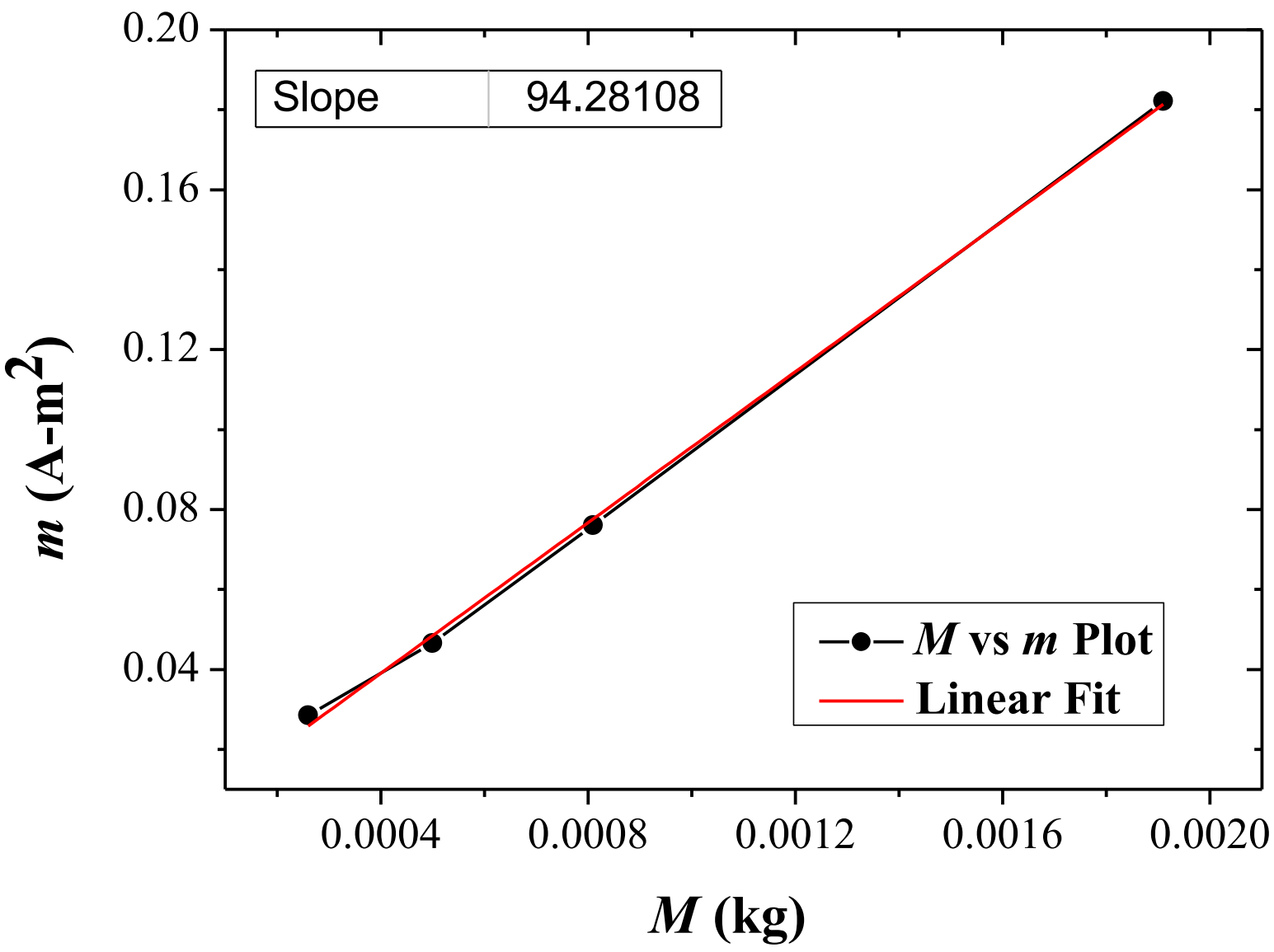

Fig. 2: Plot of mass (*M*) versus magnetic moment (*m*) of the ball magnets.

**II.** *Measurement of the terminal velocities of the ball magnets and hence determination of the acceleration due to gravity*

We have attempted to focus on the determining acceleration due to gravity based on the measurement of the magnetic moment ($m$) of each ball magnet and corresponding terminal velocity ($v_T$) during the falling motion through a metallic and non-ferromagnetic tube. Based on the Eqs. (3-6) we have finally led to the expression of g in Eq. (6), which includes $m$ and $v_T$. As the values of $m$ are already determined following method I, we need to find the terminal velocity of each ball. To this aim, we have used a copper pipe of inner and outer diameter 9.56 mm and 15.86 mm, respectively, and length 101.0 cm. The pipe is kept vertically allowing a particular ball magnet to fall through it. We have placed two identical smartphones (Samsung galaxy A51) near the pipe at the two positions (40.0 cm and 93.1 cm away from the top of the pipe) of the pipe with the phyphox magnetometer mode 'on' in those two smartphones (See Fig. 3a). We have examined that the ball magnet attains terminal velocity after traversing a minimal distance when the magnet is released into the pipe. The passage of the magnet through the pipe is monitored by using the magnetic sensors via the magnetometer of the phyphox app.

Initially, we have ensured that the copper tube was kept vertical to allow the ball magnet to move freely without any collisions. During the release of the ball magnet extra care has been taken to ensure the absence of rolling motion of the ball magnet. We have followed the same process of marking the north pole of the ball magnet with black colour, released the ball holding the north pole at the upper side and checked which part of it was at the top side when collecting the ball magnet at the bottom of the pipe. Moreover, we performed the method repeatedly and verified that, in almost all the turns, the north pole remained on the top side. At the same time, we can state that the probability of the unbalanced collision of the ball magnet and the wall of the long metal tube is less due to the strong opposing force evolved as a contribution of electromagnetic damping obeying Lenz's law. When the ball magnet falls through the pipe and passes near the smartphones, a sharp peak appears in the variation profile of the magnetic field shown in the magnetometer waveforms.

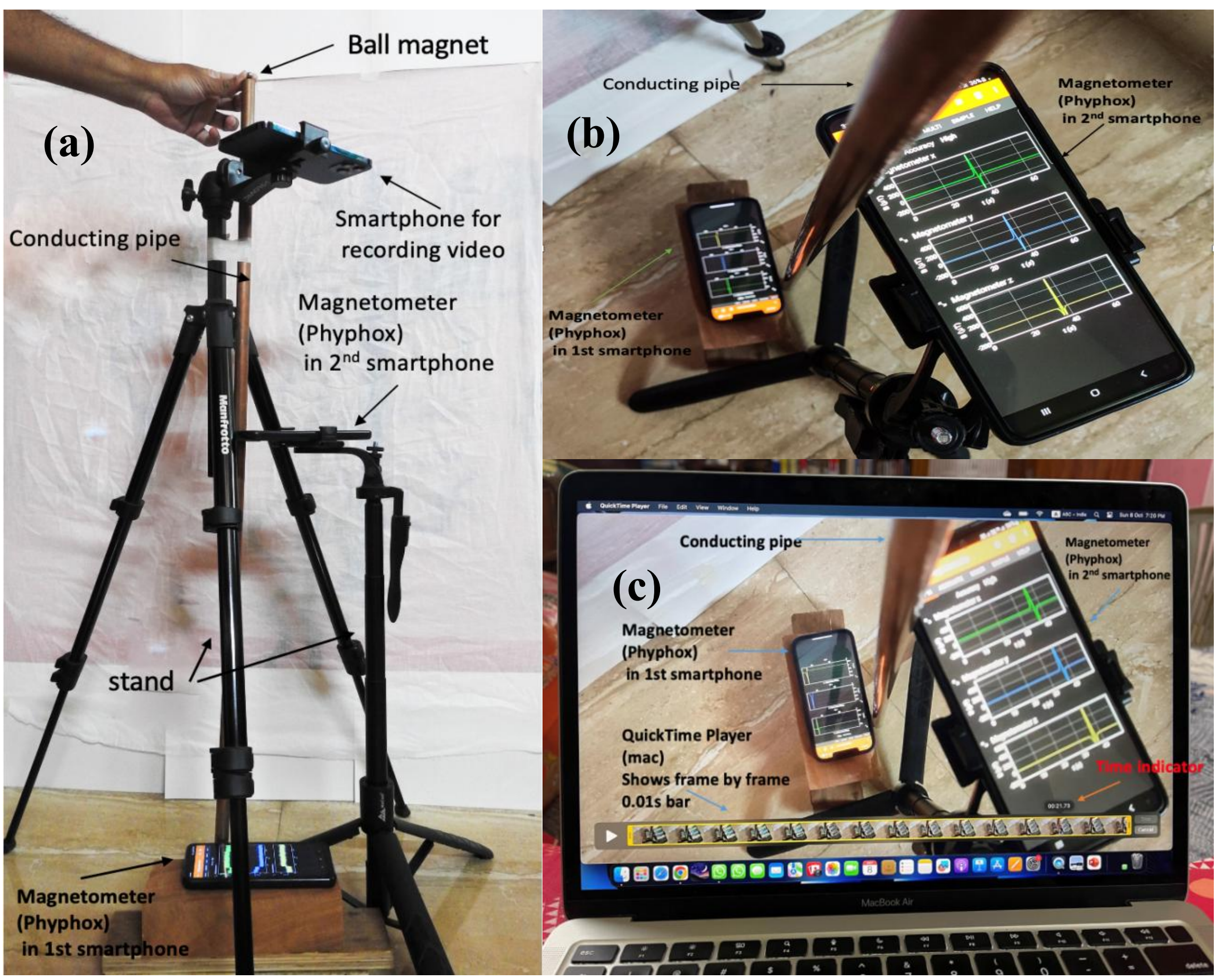

Fig. 3: (a) Experimental setup for measuring the terminal velocity of the ball magnet; (b) Two smartphones located at the two positions of the pipe are set for monitoring the passage of the magnet inside the pipe; (c) Measurement of the time interval between the two said positions for the motion of the magnet analyzing the recorded video with the help of a video player.

To compute the terminal velocity attained by the ball magnet, we need the time interval for the falling ball magnet corresponding to the occurrence of peaks in the two smartphones located at the two positions of the pipe (See Fig. 3b). It is note that the synchronization of time for the two smartphones is not necessary in view of measurement. A significant error may be introduced due to a little time difference in synchronising time for the two smartphones. A short video of the whole system along with the magnetometer waveforms in the two smartphones is recorded by another smartphone (iPhone 12 Pro Max). The frame rate of the recorder video app is 60 frames per second. By transferring the recorded video file to a MacBook Pro we have measured the said time interval with significant accuracy (minimum 0.01s) following the peaks in two different smartphones by using the QuickTime Player, the inbuilt player of the MacBook used (See Fig.

3c). When the recorded video is played using QuickTime Player and the key combination (command+T) is pressed, a millisecond bar appears. We play the video and distinguishably measure the time of showing peak by a smartphone for a particular frame. This allows us to accurately measure the time interval between two specific frames corresponding to the peaks. We easily find the terminal velocity using this time measurement technique. We repeat the experiment to have another set of data for the ball magnet with a lowering of the position of the smartphone (top one) by the length of 10.1 cm. The whole process is repeated for all other ball magnets.

Table 3: Table for the measurement of terminal velocity and there by '*g*'

| Ball magnet No. | Mean value of *m* (A-m$^2$) | Distance between two smartphones *x* (m) | Time interval between two peaks (s) | Terminal velocity, $v_T$ (ms$^{-1}$) | Acceleration due to gravity, *g* (ms$^{-2}$) | Mean value of *g* (ms$^{-2}$) |
|---|---|---|---|---|---|---|
| 1 | 0.0285 | 0.531 | 1.52 | 0.3493 | 9.77 | 9.73 ±0.46±0.01 |
| | | | 1.52 | 0.3493 | 9.77 | |
| | | | 1.53 | 0.3471 | 9.71 | |
| | | | 1.53 | 0.3471 | 9.71 | |
| | | | 1.53 | 0.3471 | 9.71 | |
| | | 0.430 | 1.24 | 0.3468 | 9.70 | |
| | | | 1.23 | 0.3496 | 9.78 | |
| | | | 1.23 | 0.3496 | 9.78 | |
| | | | 1.24 | 0.3468 | 9.70 | |
| | | | 1.24 | 0.3468 | 9.70 | |
| 2 | 0.0466 | 0.531 | 2.12 | 0.2505 | 9.81 | 9.79 ±0.36±0.02 |
| | | | 2.11 | 0.2517 | 9.83 | |
| | | | 2.13 | 0.2493 | 9.74 | |
| | | | 2.11 | 0.2517 | 9.83 | |
| | | | 2.12 | 0.2505 | 9.79 | |
| | | 0.430 | 1.72 | 0.2500 | 9.77 | |
| | | | 1.71 | 0.2515 | 9.83 | |
| | | | 1.71 | 0.2515 | 9.83 | |
| | | | 1.72 | 0.2500 | 9.77 | |
| | | | 1.73 | 0.2486 | 9.71 | |

| 3 | 0.0760 | 0.531 | 3.49 | 0.1521 | 9.78 | 9.78 ±0.31±0.02 |
|---|---|---|---|---|---|---|
| | | | 3.48 | 0.1526 | 9.81 | |
| | | | 3.49 | 0.1521 | 9.78 | |
| | | | 3.48 | 0.1526 | 9.81 | |
| | | | 3.49 | 0.1521 | 9.78 | |
| | | 0.430 | 2.83 | 0.1519 | 9.77 | |
| | | | 2.82 | 0.1525 | 9.80 | |
| | | | 2.83 | 0.1519 | 9.77 | |
| | | | 2.84 | 0.1514 | 9.73 | |
| | | | 2.83 | 0.1519 | 9.77 | |
| 4 | 0.1821 | 0.531 | 8.51 | 0.0624 | 9.76 | 9.78 ±0.24±0.02 |
| | | | 8.49 | 0.0625 | 9.78 | |
| | | | 8.47 | 0.0627 | 9.81 | |
| | | | 8.49 | 0.0625 | 9.78 | |
| | | | 8.48 | 0.0626 | 9.79 | |
| | | 0.430 | 6.88 | 0.0625 | 9.78 | |
| | | | 6.87 | 0.0626 | 9.79 | |
| | | | 6.89 | 0.0624 | 9.76 | |
| | | | 6.88 | 0.0625 | 9.78 | |
| | | | 6.87 | 0.0626 | 9.79 | |

Finally, we have determined the value of g = 9.77 ± 0.34 ± 0.018 $ms^{-2}$. We have repeated the experiment several times for each observation and calculated the uncertainty/error in the measurement. Errors are given in terms of the instrumental uncertainty (± 0.35) and the statistical error (± 0.005). The experimentally obtained value differs from the literature value by a small amount.

**Conclusion:**

In fact, smartphones with their built-in sensors, offer a versatile and practical platform for conducting physics experiments in educational settings, making complex concepts more tangible and interesting to students. In this work, we have shown the use of technology, easily available in smartphones, with neodymium ball magnets of different sizes and a non-ferromagnetic conducting pipe for measuring the acceleration due to gravity ($g$) accurately. In summary, this experiment leverages the magnetic properties of neodymium ball magnets to determine the value of g in two steps. The first step involves the computation of the magnetic moment of a magnet by measuring the time period of a torsional pendulum made by a magnet. The second step deals with the calculation of the value of $g$ by measuring the terminal velocity of a magnet when falling through the conducting pipe. The internal medium enclosed within the pipe can be envisioned as a high viscous medium due to electromagnetic damping. Finally, the work highlights that the combined use of smartphone sensors and video analysis enhances the precision of the measurements.

## Acknowledgement:

We gratefully acknowledge Dr Debapriyo Syam for stimulating discussion and his contribution to the final preparation of the manuscript. We also acknowledge Mr Papun Mondal for his cooperation during data taking.

## References:

[1] Wright, K. 2020 *Smartphone Physics on the Rise*, Physics 13, 68

[2] Stampfer, C.; Heinke, H.; Staacks, S. 2020 *A lab in the pocket*, Nat. Rev. Mater. 5, 169–170

[3] Vieyra R, Vieyra C, Jeanjacuot P, Marti A C 2015 *Turn your smartphone into a science laboratory*, The Sci. Teach 82(9), 32-40

[4] Kuhn J and Vogt P 2012 *Analyzing free fall with a smartphone acceleration sensor*, Phys. Teach. 50, 182-183

[5] Pili U and Violanda R 2018 *A simple pendulum-based measurement of g with a smartphone light sensor*, Phys. Educ. 53, 043001

[6] Silva-Alé J A 2021 *Determination of gravity acceleration with smartphone ambient light sensor*, Phys. Teach. 59, 218-219

[7] Kim J, Bouman L, Cayruth F, Elliott C, Francis B, Gogo E, Hyman C, Marshall A, Masters J, Olano W, Paone A, Patel K, Richards L, Sbardella S, Snider A, Trinh B, Umari F, and Wilks H 2020 *A measurement of gravitational acceleration using a metal ball, a ruler, and a smartphone* Phys. Teach. 58, 192–194.

[8] Deng X, Zhang J, Chen Q, Zhang J, and Zhuang W 2021 *Measurement of g using a pendulum and a smartphone proximity sensor*, Phys. Teach. 59, 584-585.

[9] Anni M 2021 *Quantitative comparison between the smartphone based experiments for the gravity acceleration measurement at home*, Educ. Sci. 11, 493 (1-20)

[10] Zhang J, Zhang J, Chen Q, Deng X, and Zhuang W 2023 *Measurement of g using a steel ball and a smartphone acoustic stopwatch*, Phys. Teach. 61, 74–75.

[11] Marín-Sepulveda C F, Castro-Palacio J C, Giménez M H, and Monsoriu J A 2023 *Acoustic determination of g by tracking a free falling body using a smartphone as a 'sonar'*, Phys. Educ. 58, 035011

[12] Pal Sanjoy K., Sarkar Soumen, and Panchadhyayee Pradipta 2024*Determination of the magnetic moment of a magnet by letting it fall through a conducting pipe*, Phys. Educ. 59**,** 015022